\documentclass[twocolumn,aps,amsmath,showpacs,pre,floatfix]{revtex4}
\usepackage{graphicx}
\usepackage{amssymb}

\usepackage{color}
\begin{document}
\title{Grand-canonical solution of semi-flexible self-avoiding trails on the Bethe lattice}
\author{W. G. Dantas}
\email{wgdantas@id.uff.br}
\affiliation{Departamento de Ci\^encias Exatas, Universidade Federal
Fluminense, Volta Redonda, RJ 27255-125, Brazil}
\author{Tiago J. Oliveira}
\email{tiago@ufv.br}

\affiliation{Departamento de F\'isica, Universidade Federal de Vi\c
  cosa, 36570-900, Vi\c cosa, MG, Brazil} 
\author{J\"urgen F. Stilck}
\email{jstilck@id.uff.br}
\affiliation{Instituto de F\'{\i}sica, Universidade Federal
  Fluminense, Av. Litor\^anea s/n, 24210-346, Niter\'oi, RJ, Brazil} 
\author{Thomas Prellberg}
\email{t.prellberg@qmul.ac.uk}
\affiliation{School of Mathematical Sciences, Queen Mary 
University of London, London E1 4NS, UK}


\begin{abstract}
We consider a model of semi-flexible interacting self-avoiding trails (sISAT's) on a lattice, where the walks are constrained to visit each lattice edge at most once. Such models have been studied as an alternative to the self-attracting self-avoiding walks (SASAW) to investigate the collapse transition of polymers, with the attractive interactions being on site, as opposed to nearest-neighbor interactions in SASAW. The grand-canonical version of the sISAT model is solved on a four-coordinated Bethe lattice, and four phases appear: non-polymerized ({\bf NP}), regular polymerized ({\bf P}), dense polymerized ({\bf DP}) and anisotropic nematic ({\bf AN}), the last one present in the phase diagram only for sufficiently stiff chains. The last two phases are dense, in the sense that all lattice sites are visited once in {\bf AN} phase and twice in {\bf DP} phase. In general, critical {\bf NP}-{\bf P} and {\bf DP}-{\bf P} transition surfaces meet with a {\bf NP}-{\bf DP} coexistence surface at a line of bicritical points. The region in which the {\bf AN} phase is stable is limited by a \textit{discontinuous critical} transition to the {\bf P} phase, and we study this somewhat unusual transition in some detail. In the limit of rods, where the chains are totally rigid, the {\bf P} phase is absent and the three coexistence lines ({\bf NP}-{\bf AN}, {\bf AN}-{\bf DP}, and {\bf NP}-{\bf DP}) meet at a triple point, which is the endpoint of the bicritical line.
\end{abstract}


\maketitle

\section{Introduction}
\label{intro}
In the most studied lattice model for the collapse transition (also called coil-globule transition) of polymers \cite{f66}, the chains are represented by self-avoiding walks, so that the bonds of the chain are placed on lattice edges and the monomers are located on the sites. An attractive interaction between monomers on nearest-neighbor sites which are not linked by bonds is added. The competition between the repulsive excluded-volume interactions and the attractive interactions leads to a change in the polymerization transition of SASAW (self-attracting self-avoiding walks) in a grand-canonical formalism. Experimentally, as the temperature of a polymer solution is lowered, the chain changes its configuration from extended (coil) to collapsed (globule), as the temperature crosses a particular value, called the $\Theta$-point. For weak attraction, the transition between a non-polymerized and a polymerized phase in the monomer fugacity-temperature plane is continuous, becoming discontinuous as the attraction is increased, so that the collapse transition is a tricritical point in this model. The nature of this transition was studied through a mapping of the polymer model onto a ferromagnetic $O(n)$ model in the limit $n \to 0$, due to de Gennes \cite{dg75,dg79}. The contributions to the high-temperature series expansion of the magnetic model are represented by self-avoiding walks on the lattice.  Mean field tricritical exponents are found in three dimensions, with logarithmic corrections.

In two dimensions, non-classical exponents are expected, and a major result is due to Duplantier and Saleur (DS), who managed to derive the exact tricritical exponents for the SASAW model on a honeycomb lattice \cite{ds87}. The proposal of this model, which requires some fine tuning to allow it to be solved, as a generic result for the collapse transition has been discussed in the literature soon after its proposal, and lead to numerical results which seem to support its robustness \cite{ss88,ds89,po94,gh95,nt14}. Another aspect of the problem are the phase diagrams of the variety of models related to the problem of the collapse transition. Even a slight change in the SASAW model, if we assume that the attractive interactions are between polymer bonds on opposite sides of elementary squares of the square lattice, leads to a phase diagram which is different from the one found when the interactions are between monomers on nearest neighbor (NN) sites. In this model, an additional polymerized phase appears, besides the non-polymerized and the regular polymerized phases, and the critical polymerization line ends at a critical endpoint \cite{jurgen,foster07}. 

The self-avoidance constraint may also be changed, allowing for more than one monomer at the same site, but still restricting the number of polymer bonds on each lattice edge to at most one \cite{mm75}. This generalization of SAW's, usually called trails, has the distinctive feature that the interactions are now between monomers {\em at the same site}. On two-dimensional lattices, the trails may collide or cross at each site, and in the original model, which we will call the ISAT (interacting self-avoiding trails) model, the statistical weights of both configurations are the same. If, the trails are not allowed to cross themselves, so that only collisions of the trails on sites exist, we have a model called VISAW (vertex interacting self-avoiding walks), which was exactly solved by Bl\"ote and Nienhuis (BN) \cite{bn89} and has critical exponents for the collapse transition distinct from the ones found in the DS model. There has been much discussion in the literature on which of the two distinct sets of exponents (DS or BN) is the generic result for the collapse transition of polymers. The BN exponents seem to be difficult to find in simulations, since numerical results for the exponents of the BN model seem to be closer to the DS values \cite{b13}. The inclusion of stiffness in the VISAW model, associating a bending energy to elementary bends of the chain, leads to an even richer phase diagram \cite{v15}, with the tricritical points from both integrable models (DS and BN) residing on a multicritical line. The robustness of the DS exponents has also been discussed in a recent paper \cite{n15}, where it has been shown that if crossings of the trails are included in the BN model, so that the lattice is no longer planar, the universality class is changed.

Although of course the analytic results mentioned above are of inestimable value, details of the phase diagrams of the different models are not always easily found with these techniques, and this is also true for simulations, in particular in the grand-canonical formalism, where the nature of the collapse transition was recognized to be tricritical by de Gennes. It is, therefore, interesting to study these phase diagrams with approximate analytic tools, among them solutions on hierarchical lattices such as the Bethe and the Husimi lattice. Indeed, the ISAT model was recently studied on a four-coordinated Husimi lattice built with squares and on a four and six-coordinated Bethe lattice \cite{tj16}. Rich phase diagrams were found in these studies, with the coil-globule transition for four-coordinated cases (which are mean-field approximations for the square lattice) being associated with a bicritical point. Such behavior was confirmed in a recent study by Pretti \cite{p16}, where the ISAT model was generalized by including an attractive interaction between NN monomers on single occupied sites not linked by a polymer bond and solved on Bethe and Husimi lattices with $q=4$. The VISAW model (when the crossings are forbidden and NN interactions vanish), the SASAW (when crossings and collisions are not allowed), the model by Wu and Bradley \cite{wb90} (when collisions and crossings have the same weight) and the simple ISAT model (when the NN interaction vanishes) are recovered as particular cases.

Given the relevance of the semi-flexible extension of the VISAW model \cite{v15} in the discussion about the differences between the DS and the BN critical behaviors for the collapse transition, we investigate another generalization of trail models by including an energy associated to elementary bends. This is done here for Bethe lattice, with different statistical weights associated to crossings and collisions, so that semi-flexible VISAW and semi-flexible ISAT are obtained as particular cases of our model. It is shown that the inclusion of semi-flexibility does not change the nature of the collapse transition when compared with the flexible ISAT model studied before \cite{tj16}, but an additional polymerized phase appears inside the regular polymerized phase, which is both dense and nematic, since all lattice sites are visited and all bonds are in the same direction. For SAW, the semi-flexible and the self-attracting cases were studied on the Bethe lattice \cite{bs93}. When both effects are present, studies on Bethe and Husimi lattices show the appearance of a second polymerized phase, which is dense and anisotropic in the sense that bonds in one particular direction are favored \cite{l98,p02}. The nature of the collapse transition is changed when the stiffness is sufficiently high, so that in this case also the appearance of a second polymerized phase signals the change of the nature of the collapse transition.

The model we study is defined in more detail in Section \ref{defmod} and its solution on a Bethe lattice is presented in Section \ref{sbl}. Final discussions and the conclusion may be found in Section \ref{conc}.

\section{Definition of the model} 
\label{defmod}

\begin{figure}[t]
\centering
\includegraphics[width=7.0cm]{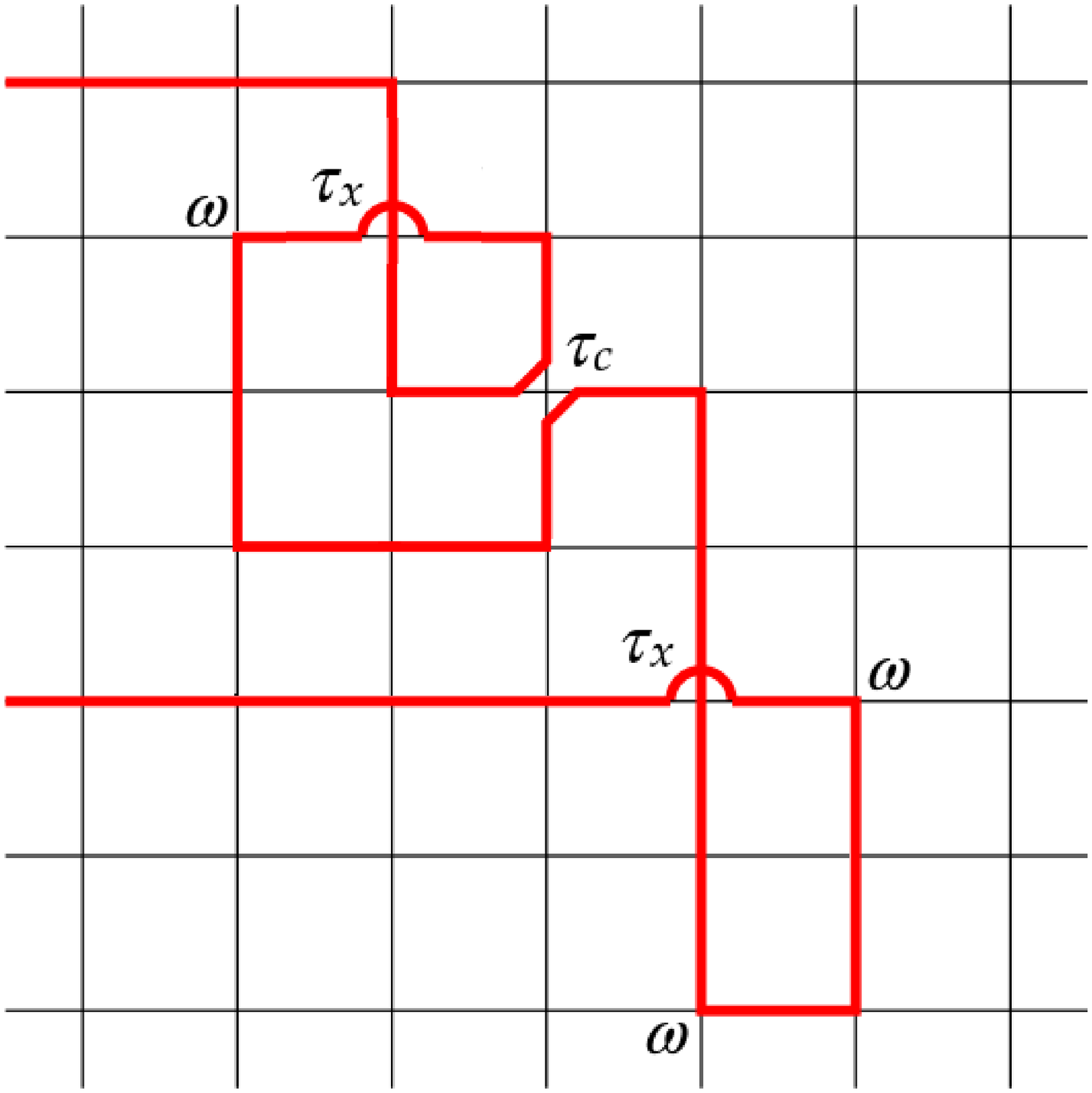}
\caption{Illustration of a self-avoiding trail on a square lattice. Collisions and crossings are indicated by $\tau_c$ and $\tau_x$, respectively, while bends at sites with two incoming bonds are indicated by $\omega$.}
\label{FigModRQ}
\end{figure}

We consider a semi-flexible generalized self-avoiding trail (sISAT) model. In this model, each lattice edge can be occupied by at most one polymer bond, which has an activity $z=\exp(\beta\mu)$, where $\beta=1/(k_BT)$ and $\mu$ is the chemical potential of a bond. The bonds connect monomers, which are placed on the sites of the lattice. For a lattice with coordination number $q$, each lattice site can be occupied by up to $q/2$ (for even $q$) and $(q-1)/2$ (for odd $q$) monomers. In two dimensions, walks meeting at a lattice site may either cross or collide, as is apparent in the generalized ISAT model on a square lattice depicted in Fig.~\ref{FigModRQ}. In higher dimensions, however, the distinction between collisions and crossings is no longer clear. We will restrict our attention to lattices with $q=4$ here, whose solutions may be compared with results for the square lattice. Statistical weights $\tau_x$ and $\tau_c$ will be associated for each crossing and collision of the chains, respectively. Note that when $\tau_x=\tau_c$ we recover the classical ISAT model, while in the case $\tau_x=0$ (crossings forbidden) the VISAW model is obtained. In order to analyze the effects of the polymer stiffness in such models, an additional weight $\omega=\exp(-\beta\epsilon_b)$ is introduced, associated to one polymer bend. If the energy $\epsilon_b$ associated to an elementary bend of the trails is positive ($\omega<1$), we say that the walks are semi-flexible. This is also illustrated in Fig.~\ref{FigModRQ}. The grand-canonical partition function of the model is given by
\begin{equation}
 Y = \sum z^{N} \tau_c^{N_{c}} \tau_x^{N_{x}} \omega^{N_{b}},
 \label{eqPartFunc}
\end{equation}
where $N$, $N_c$ and $N_x$ are the numbers of bonds, collisions and crossings in the system, respectively. $N_b$ is the number of bends in sites with a single monomer, since the bends in double visited sites are accounted for in the weight $\tau_c$, so that $\tau_c=\omega^2 \tau^*$, where $\tau^*$ is the weight of the collision itself, i. e., of the \textit{on-site} monomer-monomer interaction at colliding sites. We will mostly restrict ourselves in the discussions to $\tau_x=\tau^*=\exp(-\beta\epsilon)$, so that the monomer-monomer interaction energy $\epsilon$ is the same for crossings and collisions, but it is easy to consider $\epsilon_c \neq \epsilon_x$, and this will be done in part of the comments below. The sum in Eq.~(\ref{eqPartFunc}) is over all allowed configurations of the walks on the lattice we are considering, which will be the four coordinated Bethe lattice here. 

\section{Solution on the Bethe lattice}
\label{sbl}
\begin{figure}[t]
\centering
\includegraphics[width=7.0cm]{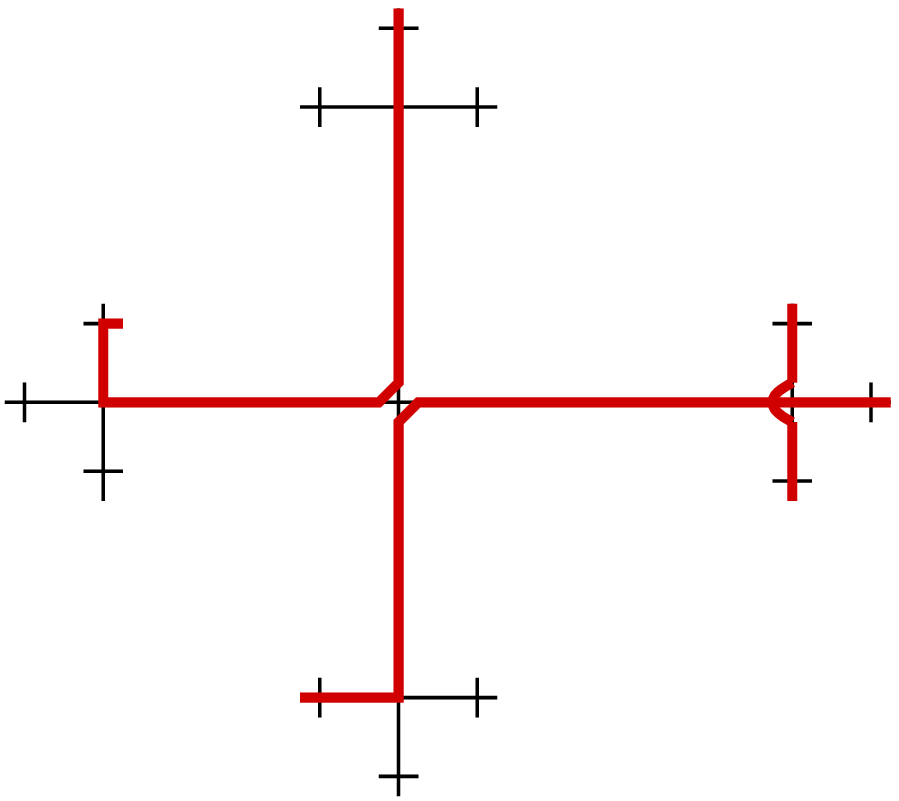}
\caption{A configuration of three chains placed on a Cayley tree with four generations of sites. The statistical weight of this configuration is $z^{16}\omega^3\tau_c\tau_x$.}
\label{bl}
\end{figure}

We solve the model on a four coordinated Bethe lattice, which corresponds to the core of a Cayley tree as shown in Fig.~\ref{bl}. The extremal monomers of each chain are placed on the surface of the tree. One possible configuration of three chains on a Cayley tree with four generations of sites is shown in Fig.~\ref{bl}.

To solve the model on the Bethe lattice  we consider sub-trees, defining partial partition functions (ppf's) for them for a fixed configuration of the root edge \cite{b82}. For the Bethe lattice, usually only two root configurations are needed, corresponding to the possibilities of empty or occupied (by chain bonds) root edges. However, as discussed above, in the semi-flexible case one may expect the appearance of anisotropic phases, which display orientational (nematic) ordering, so that bonds in one or more directions are favored. The study of models which present nematic ordering on such hierarchical lattices presents some difficulties, particularly for $q>4$. One way to avoid them is to solve these models on other treelike lattices for which the exact solution is the Bethe approximation on regular lattices with the same coordination number. One of such lattices is the random locally treelike layered (RLTL) lattice introduced by Dhar, Rajesh and Stilck \cite{drs11} to study nematic ordering of monodispersed rigid rods. Here we will follow a simpler option, assuming that at each site of the $q=4$ Bethe lattice two incoming bonds are in one direction and the two remaining ones are in a perpendicular direction. Actually, one should keep in mind that in the thermodynamic limit this lattice is effectively infinite dimensional, as was shown by Baxter \cite{b82}. Thus, for example, to correctly measure the Euclidean distance between two sites on an even-coordinated Bethe lattice, one may embed it in a hypercubic lattice whose dimension increases with the number of generations \cite{sca00}. Therefore, we will define partial partition functions for four root configurations: $g_{0,1}$ for a root edge in direction $1$ not occupied by a bond, $g_{0,2}$ for a empty root edge in direction $2$, and $g_{1,1}$ and $g_{1,2}$ for subtrees with occupied root edges in directions $1$ and $2$, respectively.

We now may obtain recursion relations for the ppf's of a sub-tree with an additional generation, considering the operation of attaching a new root site and edge to three sub-trees of the preceding generation. The results are:
\begin{subequations}
\begin{eqnarray}
g'_{0,1} &=& g_{0,1}g_{0,2}^2+g_{0,1}g_{1,2}^2+2\omega g_{1,1}g_{1,2}g_{0,2}, \\
g'_{0,2} &=& g_{0,1}^2g_{0,2}+g_{1,1}^2g_{0,2}+2\omega g_{0,1}g_{1,1}g_{1,2}, \\
g'_{1,1} &=& z_1[g_{1,1}g_{0,2}^2+2\omega g_{0,1}g_{0,2}g_{1,2}+\tau g_{1,1}g_{1,2}^2], \\
g'_{1,2} &=& z_2[g_{0,1}^2g_{1,2}+2\omega g_{0,1}g_{1,1}g_{0,2}+\tau g_{1,1}^2g_{1,2}],
\end{eqnarray}
\label{rrgb}
\end{subequations}
where $\tau \equiv \tau_x+2\tau_c$ is the only combination of the weights of double occupied sites that appears in the Bethe lattice solution; this will change if longer range correlations are taken into account, such as on the Husimi lattice. We note that we include the possibility of bonds in the two directions having different activities, although we will discuss the thermodynamic behavior of the model only for $z_1=z_2=z$.

Here, $g_{i,j}$ and $g'_{i,j}$ are ppf's of sub-trees with $M$ and $M+1$ generations, respectively. Usually, the ppf's diverge in the thermodynamic limit (when $M \rightarrow \infty$). Thus, it is suitable to define the ratios: 
\begin{subequations}
\begin{eqnarray}
R_1 &=& \frac{g_{1,1}}{g_{0,1}}, \\
R_2 &=& \frac{g_{1,2}}{g_{0,2}},
\end{eqnarray}
\end{subequations}
which should remain finite for non-dense phases, where a finite fraction of the lattice sites is empty. The recursion relations for these ratios are:
\begin{subequations}
\begin{eqnarray}
R'_1 &=& z_1\frac{R_1+2\omega R_2+\tau R_1R_2^2}{1+R_2^2+2\omega R_1R_2},\\
R'_2 &=& z_2\frac{R_2+2\omega R_1+\tau R_1^2R_2}{1+R_1^2+2\omega R_1R_2}.
\end{eqnarray}
\label{rrrb}
\end{subequations}

We find four distinct fixed points for these recursion relations when $z_1=z_2=z$, which are stable in distinct regions of the parameter space $(z,\omega,\tau)$. They are:
\begin{itemize}
\item
A non-polymerized ({\bf NP}) phase: $R_1=0$, $R_2=0$.
\item
A regular polymerized ({\bf P}) phase: $R_1=R_2 \ne 0$ and finite.
\item 
A dense polymerized ({\bf DP}) phase: $R_1=R_2 \to \infty$.
\item
A dense anisotropic and nematic ({\bf AN}) phase: $R_1 \to \infty$ and $R_2 \to 0$ or $R_1 \to 0$ and $R_2 \to \infty$.
\end{itemize}
In the dense phases, the edges corresponding to the direction of the ratio which diverges are all occupied by bonds. It is useful to define the reciprocal ratios $S_i=1/R_i$ to study the fixed points which are associated to these phases. For the {\bf DP} phase we may rewrite the recursion relation Eqs. (\ref{rrrb}) as:
\begin{subequations}
\begin{eqnarray}
S'_1 &=& \frac{1}{z}\frac{S_1S_2^2+S_1+2\omega S_2}{S_2^2+2\omega S_1S_2+\tau},\\
S'_2 &=& \frac{1}{z}\frac{S_1^2S_2+S_2+2\omega S_1}{S_1^2+2\omega S_1S_2+\tau}.
\label{rrdp}
\end{eqnarray}
\end{subequations}
So that the {\bf DP} fixed point is now located at the origin $S_1=0,\;\;S_2=0$. For the anisotropic {\bf AN} phase, there are two equivalent fixed points where the chains occupy bonds in one of the two directions. If we consider the fixed point with bonds in the 1 direction, the recursion relations may be written in terms of the variables $S_1$ and $R_2$, so that the fixed point again is located at the origin. The result is:
\begin{subequations}
\begin{eqnarray}
S'_1 &=& \frac{1}{z}\frac{S_1+S_1R_2^2+2\omega R_2}{1+2\omega S_1R_2+\tau R_2^2},\\
R'_2 &=& z\frac{S_1^2R_2+2\omega S_1+\tau R_2}{1+S_1^2+2\omega S_1R_2}.
\label{rrdn}
\end{eqnarray}
\end{subequations}
We notice that the product $P=R_1R_2$ attains a finite value at the {\bf AN} fixed point, given by:
\begin{equation}
P=\frac{z^2\tau-1+\sqrt{(z^2\tau-1)^2+16 z^2\omega^2}}{4\omega}
\label{P}
\end{equation}

The region of the parameter space where each fixed point is stable may be found by studying the eigenvalues of the Jacobian of the recursion relations:
\begin{equation}
J_{i,j}=\frac{\partial Q'_i}{\partial Q_j},
\end{equation}
where the $Q$'s are the appropriate ratios in each case. In the {\bf NP} fixed point $R_1=R_2=0$, the secular equation of the recursion relation (\ref{rrrb}) is 
\begin{equation}
(z-\lambda)^2-4z^2\omega^2=0, 
\end{equation}
so that this fixed point is stable for:
\begin{equation}
 z \le 1/(1+2\omega). 
\label{slnp}
\end{equation}
The secular equation associated to the {\bf DP} fixed point $S_1=S_2=0$ is:
\begin{equation}
\left(\frac{1}{z\tau}-\lambda\right)^2-\left(\frac{2\omega}{z\tau}\right)^2=0,
\end{equation}
and thus the region where this phase is stable will be:
\begin{equation}
\tau \ge \frac{1+2\omega}{z}
\label{sldp}
\end{equation}
The secular equation for the {\bf AN} fixed point $S_1=0$, $R_1=0$ is:
\begin{equation}
\lambda^2-\left(\frac{1}{z}+z\tau\right)\lambda+\tau-4\omega^2=0.
\label{sedn}
\end{equation}
From which it follows that this fixed point is stable if 
\begin{equation}
\tau \le \frac{1}{z}-\frac{4\omega^2}{z-1}. 
\label{sldn}
\end{equation}
Since $\tau$ (as well as $z$ and $\omega$) are non-negative, the {\bf AN} phase is stable for allowed values of $\tau$ if $z \geq 1/(1-4 \omega^2)$, so that it exists only when $\omega < 1/2$. This is expected since the chains should be sufficiently stiff to generate nematic order.

For the isotropic polymerized fixed point {\bf P}, where $R_1=R_2=R$, the elements of the Jacobian are $J_{1,1}=J_{2,2}=A$ and $J_{1,2}=J_{2,1}=B$, with:
\begin{subequations}
\begin{eqnarray}
A&=&\frac{z+(\tau z-2\omega)R^2}{1+(1+2\omega)R^2}, \\
B&=&2\frac{\omega z+(\tau z-1-\omega)R^2}{1+(1+2\omega)R^2},
\end{eqnarray}
\end{subequations}
where the squared ratio of ppf's is given by:
\begin{equation}
R^2=\frac{(1+2\omega)z-1}{1+2\omega-z\tau},
\label{rfp}
\end{equation}
and thus the stability condition for this fixed point will be 
\begin{subequations}
\begin{eqnarray}
A+B &\le& 1 \\
A-B &\le& 1.
\end{eqnarray}
\label{slp}
\end{subequations}
We find that this condition is obeyed in the region of the parameter space between the surfaces $z=1/(1+2\omega)$, $z\tau=1+2\omega$, and $\tau=1/z-4\omega^2/(z-1)$, which correspond to the stability limits of the {\bf NP}, {\bf DP}, and {\bf AN} fixed points, respectively. Note that $\tau^{AN} \rightarrow \tau^{DP}$ when $\omega \rightarrow 0$, so that the {\bf P} phase disappears in this limit. As will be shown below, although the order parameter is discontinuous at the {\bf P-AN} transition, the two stability limits meet at the transition surface. For $\omega > 0$, the two critical surfaces {\bf NP-P} and {\bf DP-P} meet at the bicritical line, located at:
\begin{subequations}
\begin{eqnarray}
z&=&\frac{1}{1+2\omega} \\
\tau&=&(1+2\omega)^2.
\end{eqnarray}
\label{bcl}
\end{subequations} 

Before proceeding, we will discuss the possibility of regular polymerized phases with nematic order, with distinct and finite ratios in both directions. Summing and subtracting the fixed point equations ($R'_i=R_i$ in Eqs. (\ref{rrrb})) we obtain:
\begin{subequations}
\begin{eqnarray}
(R_1+R_2)[(2\omega-z\tau+1)P-z(1+2\omega)+1]=0, \\
(R_1-R_2)[(2\omega+z\tau-1)P-z(1-2\omega)+1]=0,
\end{eqnarray}
\end{subequations}
where we recall that $P=R_1R_2$. If the phase is polymerized and nematic, the second factors of both equations have to vanish. This leads to:
\begin{subequations}
\begin{eqnarray}
P&=&\frac{z-1}{2\omega},\\
P&=&\frac{2\omega z}{1-z\tau}.
\label{prod}
\end{eqnarray}
\end{subequations}
We notice that the conditions $z>1$ and $\tau z <1$ must be satisfied, and also the three parameters of the model are not independent in the nematic phase with finite ratios, since they are related by the equation:
\begin{equation}
\frac{z-1}{2\omega}=\frac{2\omega z}{1-z\tau},
\label{nc}
\end{equation}
which happens to be equivalent to the stability limit of the {\bf AN} phase above, Eq.~(\ref{sldn}). Thus, on this surface of the parameter space, we have a continuous set of marginally stable (with $\lambda=1$) fixed points $P=const$, which includes the regular polymerized fixed point $R_1=R_2$ and the {\bf AN} fixed point $R_1 \to \infty$, $R_2 \to 0$. Therefore, we have a discontinuous transition between these two phases, but the transition surface is not between two spinodal surfaces of the two phases which coexist. Incidentally, we notice that the value of the product of ratios in the {\bf AN} phase [Eq.~(\ref{P})] reduces to Eq.~(\ref{prod}) on the stability limit of this phase. This rather unusual feature of the {\bf AN}-{\bf P} transition, which is critical but has a discontinuous order parameter, has been discussed in the literature before. One simple situation of this kind is the one-dimensional Ising model with interactions decaying with the distance between spins as $1/r^2$ \cite{t69}. At zero field, in this model a discontinuous magnetization is found at the critical point. The one-dimensional Ising model with nearest-neighbor interactions has also been studied via exact renormalization-group transformations by Nelson and Fisher \cite{nf75}, and, although the phase transition there is degenerate, since it happens at zero temperature, it may be interpreted also as a critical discontinuous transition. The possibility of such transitions in the framework of the renormalization group was discussed in general by Fisher and Berker \cite{fb82}. This unusual critical behavior was also found in the stationary behavior of non-equilibrium models associated to the Ising model and in the threshold contact process \cite{mjo93}.

The partition function of the model on the whole Cayley tree is obtained considering the operation of connecting four subtrees to the central site. The result is:
\begin{eqnarray}
Y&=&(g_{0,1}g_{0,2})^2+(g_{1,1}g_{0,2})^2+(g_{0,1}g_{1,2})^2+ \nonumber \\
&&4\omega g_{1,1} g_{0,1} g_{1,2} g_{0,2}+\tau (g_{1,1}g_{1,2})^2,
\label{pfct}
\end{eqnarray}
where the first term corresponds to the configuration with no bond incident on the central site, the next three have two incident bonds and in the last all four edges are occupied. This expression may also be written as $Y = (g_{0,1}g_{0,2})^2 y$, with
\begin{equation}
 y=1+R_1^2+R_2^2+4\omega R_1R_2+\tau (R_1R_2)^2.
\label{pfct1}
\end{equation}
The densities of bonds in both directions, and bends (in non-colliding sites), collisions and crossings at the central site are given respectively by: 
\begin{eqnarray}
\rho_{z,1} &=& \frac{R_1^2+2\omega R_1R_2+\tau R_1^2R_2^2}{y}\\
\rho_{z,2} &=& \frac{R_2^2+2\omega R_1R_2+\tau R_1^2R_2^2}{y}\\
\rho_\omega&=&\frac{4\omega R_1R_2}{y} \\
\rho_c&=&\frac{2\tau_c R_1^2R_2^2}{y} \\
\rho_x&=&\frac{\tau_x R_1^2R_2^2}{y}
\end{eqnarray}
As discussed before, the total density of bends is $\rho_b=\rho_\omega+2\rho_c$. In the {\bf NP} phase all densities vanish. In the {\bf AN} phase with $R_1 \to \infty$ and $R_2 \to 0$, $\rho_{z,1}=1$ and all other densities vanish. In the {\bf DP} phase, $\rho_{z,1}=\rho_{z,2}=1$, $\rho_\omega=0$, $\rho_c=2\tau_x/(2\tau_z+\tau_x)$, and $\rho_x=\tau_x/(2\tau_c+\tau_x)$. Finally, in the {\bf P} phase, $\rho_{z,1}=\rho_{z,2}= [z(1+2\omega)-1]/[2z(1+2\omega)-1-\tau z^2]$, $\rho_\omega=4\omega R^2/y$, $\rho_c=2\tau_cR^4/y$, and $\rho_x=\tau_xR^4/y$, where the ratio $R$ is given by Eq.~(\ref{rfp}). 

The bulk free energy per site on the BL, which differs from the one obtained before in Eq.~(\ref{pfct}) by the contribution of the surface of the Cayley tree, may be found following an Ansatz by Gujrati \cite{g95}. The result is:
\begin{equation}
\frac{\phi_b}{k_BT}=-\lim_{M \to \infty}\frac{1}{2}\ln \frac{Y_{M+1}}{Y_M^3}.
\end{equation}
From the recursion relations (Eqs.~(\ref{rrgb})) we have that:
\begin{equation}
\frac{\phi_b}{k_BT}=-\frac{1}{2} \ln \frac{Y^\prime}{Y^3},
\label{phibi}
\end{equation}
where $Y^\prime$ is the partition function calculated with ppf's of subtrees with an additional generation with respect to the unprimed ppf's and the thermodynamic limit is implicit. In the {\bf NP} fixed point, $g_{0,1}$ and $g_{0,2}$ are dominant over the other terms in the partition function Eq.~(\ref{pfct}), so that we may rewrite Eq.~(\ref{phibi}) as:
\begin{equation}
\phi_b^{(NP)}=-\frac{k_BT}{2} \ln \frac{(g_{0,1}^\prime)^2(g_{0,2}^\prime)^2}{[g_{0,1}^2 g_{0,2}^2]^3}=0,
\label{phinp}
\end{equation}
where we have used the recursion relations in Eqs. (\ref{rrgb}). This result is consistent with the fact that this phase corresponds to an empty lattice. In the {\bf DP} phase, the last term of the partition function dominates over the others, so that:
\begin{eqnarray}
\phi_b^{(DP)}&=&-\frac{k_BT}{2}\ln \frac{\tau (g_{1,1}^\prime)^2 (g_{1,2}^\prime)^2}{(\tau g_{1,1}^2 g_{1,2}^2)^3}=-k_BT \ln(z^2\tau)= \nonumber \\
&&\epsilon-2\mu-k_BT\ln[1+\exp(-2\beta \epsilon_B)],
\label{phidp}
\end{eqnarray}
where we recall that in this phase four bonds are incident on each site. In the {\bf AN} phase, supposing that the bonds are in the 1 direction, the  second term of the partition function dominates, so that:
\begin{equation}
\phi_b^{(AN)}=-\frac{k_BT}{2}\ln \frac{(g_{1,1}^\prime)^2(g_{0,2}^\prime)^2}{(g_{1,1}^2g_{0,2}^2)^3}=-k_BT \ln z=-\mu,
\label{phian}
\end{equation}
and again this confirms that in this phase each site has two bonds in direction 1 incident on it. Finally, in the regular polymerized phase {\bf P}, where $R_1=R_2=R$, we may rewrite Eq.~(\ref{phibi}) as:
\begin{eqnarray}
\phi_b^{(P)}&=&-\frac{k_BT}{2} \ln \frac{(g_{0,1}^\prime)^2(g_{0,2}^\prime)^2}{(g_{0,1} g_{0,2})^6 y^2}= \nonumber \\
&&-k_BT \ln \frac{[\tau-(1+2\omega)^2]z^2}{1-2(1+2\omega)z+\tau z^2}.
\end{eqnarray}

\subsection{Phase diagrams}

Beyond the critical surfaces (continuous for the {\bf NP-P} transition and {\bf P-DP} and discontinuous for {\bf P-AN} transition), we note that the {\bf NP} and {\bf DP} phases coexist in the region $z<1/(1+2\omega)$ and $\tau>(1+2\omega)/z$. The discontinuous {\bf NP-DP} transition is located at the surface where the bulk free energies per site of the two phases are equal and, from Eqs.~(\ref{phinp}) and (\ref{phidp}), we find it at $\tau=1/z^2$. This surface ends at the bicritical line (Eqs.~\ref{bcl}). As expected, at all other transition surfaces the respective bulk free energies of the involved phases are also equal.

\begin{figure}[!t]
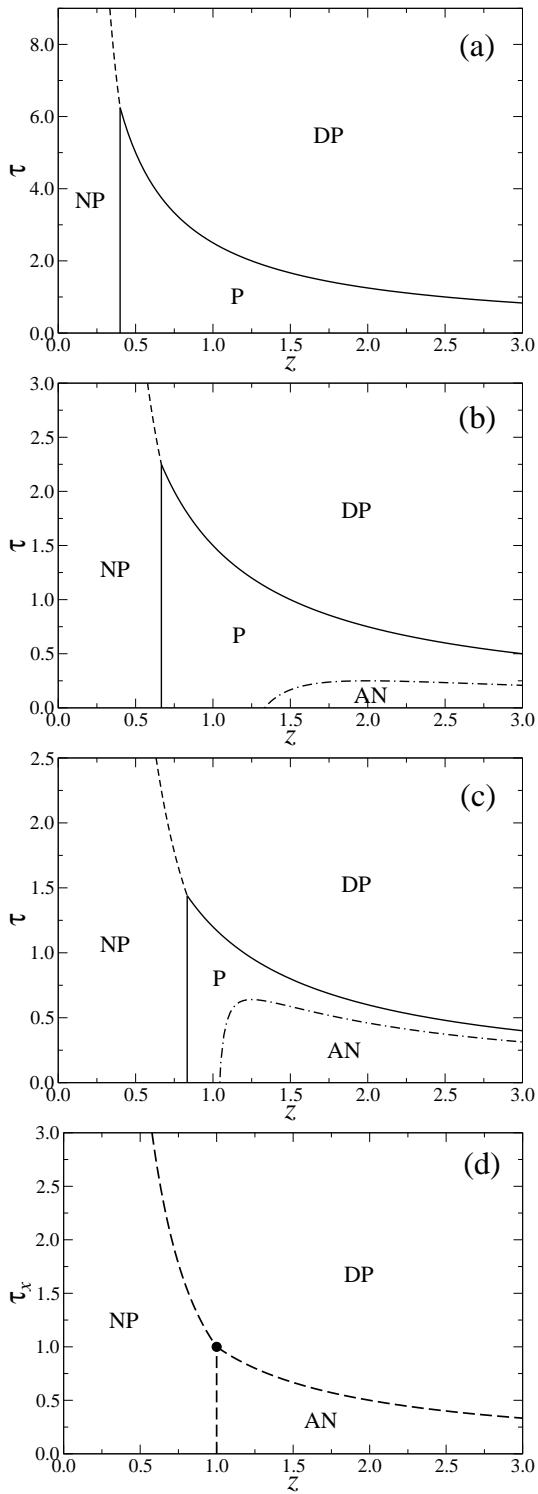

\centering
\includegraphics[width=7.cm]{Fig3a.eps}
\includegraphics[width=7.cm]{Fig3b.eps}
\includegraphics[width=7.cm]{Fig3c.eps}
\includegraphics[width=7.cm]{Fig3d.eps}
\caption{Phase diagrams for (a) $\omega=0.75$, (b) $\omega=0.25$, (c) $\omega=0.1$ and (d) $\omega=0.0$. Regular continuous transitions are shown as solid lines, discontinuous transitions are represented as dashed lines and the dash-dotted line corresponds to the critical discontinuous transitions. The dot in (d) the endpoint of the bicritical line, where the three discontinuous transition lines meet.}
\label{pd}
\end{figure}

As discussed above, for $\omega \geq 1/2$ the {\bf AN} phase is not present in the phase diagrams, see an example in Fig.~\ref{pd}(a) for $\omega=0.75$, which is qualitatively similar to the one obtained in the flexible case ($\omega=1$) \cite{tj16}. For $\omega<1/2$, the thermodynamic behavior is still the same, except for the presence of the {\bf AN} phase, as well as the critical discontinuous {\bf P-AN} surface. Diagrams for $\omega=0.25$ and $\omega = 0.1$ are shown in Figs.~\ref{pd}(b) and (c), respectively, where one sees that by decreasing $\omega$ the region occupied by the {\bf P} phase decreases, whilst that by the {\bf AN} phase increases.

Indeed, in the limit of rigid trails $\omega \to 0$ ($\epsilon_b \to \infty$), only the two dense phases appear, besides the {\bf NP} phase, since the {\bf P} phase between them is absent. The roots of the secular equation related to the {\bf AN} phase [Eq.~(\ref{sedn})] are $1/z$ and $\tau_x z$ in this case, so that the stability limit of this phase for $z>1$ is $\tau_x=1/z$, which coincides with the stability limit of the {\bf DP} phase [Eq.~(\ref{sldp})] (since there are no bends, $\tau=\tau_x$). Also, the stability limits of the {\bf AN} and {\bf NP} phases meet at $z=1$. As always, the {\bf NP} and the {\bf DP} phases coexist on the line $\tau_x=1/z^2$. The three transition lines meet at $z=\tau_x=1$, as may be seen in Eq.~(\ref{bcl}). As $\omega \to 0$, at the discontinuous {\bf AN}-{\bf NP} transition line, located at $z=1$, the {\bf NP}-{\bf P} critical surface and the {\bf P}-{\bf AN} discontinuous critical surface meet, while the {\bf DP}-{\bf P} critical surface and the {\bf P}-{\bf AN} discontinuous critical surface meet at the {\bf DP}-{\bf AN} discontinuous transition line, located at $\tau_x=1/z$. The phase diagram in this limit is shown in Fig.~\ref{pd}(d). Actually, it is quite simple to obtain the free energy of the model on the square lattice in this limit of rods, and the same phase diagram is obtained. This calculation is presented in the appendix.

\subsection{Densities}

\begin{figure}[b]
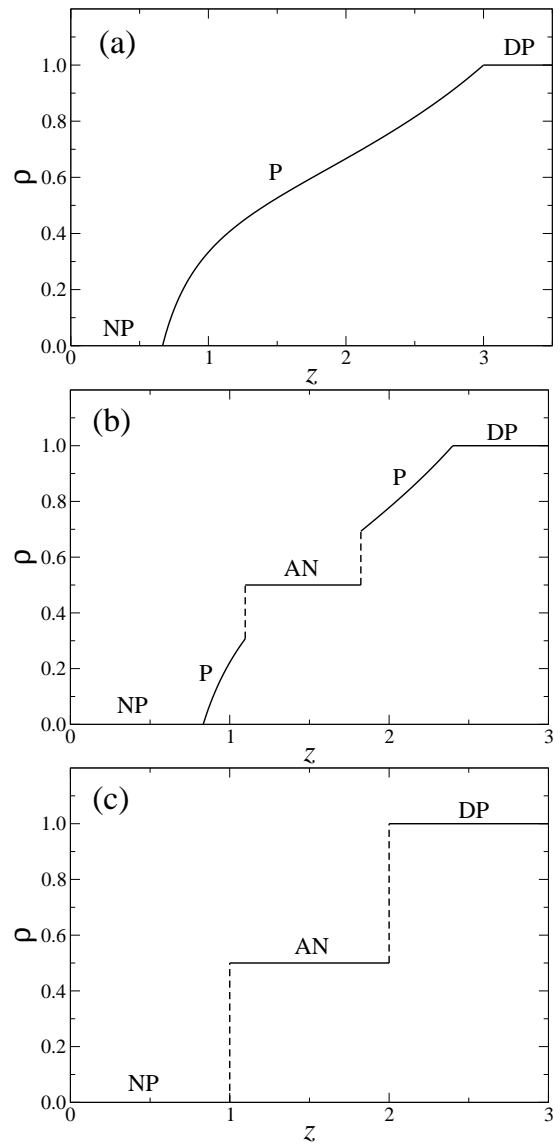

\centering
\includegraphics[width=7.20cm]{Fig4a.eps}
\includegraphics[width=7.20cm]{Fig4b.eps}
\includegraphics[width=7.20cm]{Fig4c.eps}
\caption{Total densities of bonds as functions of $z$ for $\tau=0.5$ and (a) $\omega=0.25$, (b) $\omega=0.1$ and (c) $\omega=0.0$.}
\label{rhoz}
\end{figure}

In this subsection, we investigate the behavior of the densities at the different transitions. We start noting that the total density of bonds $\rho=(\rho_{z,1}+\rho_{z,2})/2$ assumes the values: $\rho=0$ in the {\bf NP} phase, $\rho=1$ in the {\bf DP} phase, $\rho=1/2$ in the {\bf AN} phase, and $0 < \rho < 1$ in the {\bf P} phase. It may be useful to recall that the order parameter of the polymerization transition is actually $m=\rho^{1/2}$, as is shown by the mapping of the problem onto the magnetic $n$-vector model in the limit $n \to 0$ \cite{wkp80}. Moreover, we may define a nematic order parameter as:
\begin{equation}
Q=\rho_{z,1}-\rho_{z,2}=\frac{R_1^2-R_2^2}{1+R_1^2+R_2^2+4\omega R_1R_2+\tau R_1^2R_2^2},
\label{eqnop}
\end{equation}
which is $|Q|=1$ in the {\bf AN} phase and $Q=0$ otherwise, indicating that any transition to {\bf AN} phase is discontinuous. Indeed, this is confirmed in Fig.~\ref{rhoz}, which shows the variation of $\rho$ with $z$, for $\tau=0.5$ and several values of $\omega$. Close to the {\bf NP}-{\bf P} transition we have $\rho \approx (z-z_c)$, consistent with an mean field exponent $\beta=1/2$ for the order parameter.

Along the {\bf P-AN} transition surface, we have $\rho^{(P)}=(z-1)/(z-1+2\omega z)$ for the {\bf P} phase, which increases with $z$, from $\rho=2\omega/(1+2\omega)$ (at $z=1/(1-4\omega^2)$) to $\rho=1/(1+2\omega)$ (for $z \rightarrow \infty$). Between these limits, there exists a line, given by $z=1/(1-2\omega)$, where $\rho$ is continuous (i.e., $\rho^{(P)}=\rho^{(AN)}=1/2$), but $Q$ (and so, the transition) is still discontinuous. Still on the {\bf P-AN} critical surface, the infinite set of marginally stable solutions $R_1R_2=const.$ have densities $\rho_1=R_1^2/(z+R_1^2)$ and $\rho_2=(z-1)^2/((z-1)^2+4\omega^2 z R_1)$. Thereby, for fixed $\omega$ and $z$, $\rho_1$ increases ($\rho_2$ decreases) from 0 to 1 (from 1 to 0) when $R_1$ changes from $0$ to $\infty$. In both limits, we recover the {\bf AN} result and, when $\rho_1=\rho_2$ the {\bf P} phase is obtained. We note that in this phase, $\rho \neq 1/2$ for all values of $\omega$, $z$ and $R_1$, except at the line $z=1/(1-2\omega)$, where $\rho=1/2$ regardless the value of $R_1$.

\subsection{Nematic susceptibility}

\begin{figure}[t]
\centering
\includegraphics[width=7.0cm]{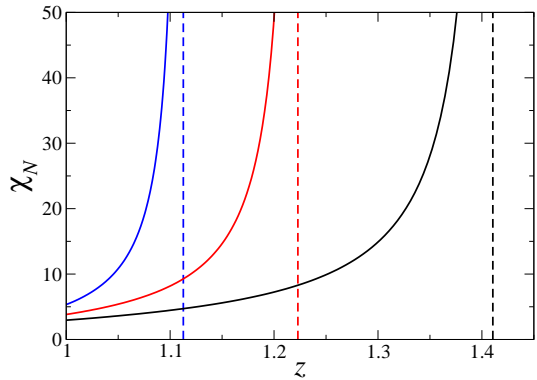}
\caption{Susceptibility $\chi_N$ in the {\bf P} phase as a function of the bond activity $z$ for $\tau=0.1$, close to the {\bf P}-{\bf AN} transition. The curves, from left to right, correspond to $\omega=0.15$ (blue), $\omega=0.20$ (red) and $\omega=0.25$ (black). Dashed lines indicate the corresponding critical activities.}
\label{chiws}
\end{figure}

Let us discuss in some more detail the {\bf AN}-{\bf P} transition by studying the behavior of the appropriate susceptibility close to it. Given the activities $z_1$ and $z_2$ of bonds in each direction, we may define
\begin{equation}
z=\frac{z_1+z_2}{2}
\end{equation}
and
\begin{equation}
{\bar z}=\frac{z_1-z_2}{2}.
\end{equation}
The activity ${\bar z}$ is the appropriate field-like variable conjugated to the nematic order parameter, so that we define the nematic susceptibility as:
\begin{equation}
\chi_N=\left(\frac{\partial Q}{\partial {\bar z}}\right)_{z,\omega,\tau},
\end{equation}
where the nematic order parameter $Q$ is defined in Eq.~(\ref{eqnop}). To obtain an expression for $\chi_N$ in the {\bf P} phase, we start with the fixed point equations which follow from the recursion relations Eqs.~(\ref{rrrb}), remembering that $z_1=z+{\bar z}$ and $z_2=z-{\bar z}$. Differentiating these equations with respect to ${\bar z}$, we obtain a system of linear equations for the derivatives of the ratios, whose solution is (for ${\bar z}=0$):
\begin{subequations}
\begin{eqnarray}
\left(\frac{\partial R_1}{\partial {\bar z}}\right)_{z,\omega,\tau}&=&\frac{F(R_1,R_2;z,\omega,\tau)}{D(R_1,R_2;z,\omega,\tau)}, \\
\left(\frac{\partial R_2}{\partial {\bar z}}\right)_{z,\omega,\tau}&=&-\frac{F(R_2,R_1;z,\omega,\tau)}{D(R_2,R_1;z,\omega,\tau)}, 
\end{eqnarray}
\end{subequations}
where we have
\begin{widetext}
\begin{eqnarray}
F(R_1,R_2;z,\omega,\tau)&=&(1-z-4\omega^2z)R_1+2\omega(1-2z)R_2+ 
(1+4\omega^2-\tau z)R_1^3+4\omega(3-2\tau z)R_1^2R_2+ \nonumber \\
&&(2+\tau+8\omega^2-3\tau z)R_1R_2^2+3\tau(1-\tau z)R_1^3R_2^2+
4\tau\omega R_1^2R_2^3+2\tau\omega R_1^4R_2,\\
D(R_1,R_2;z,\omega,\tau)&=&1+z^2-2z(1+2\omega^2z)+
[4\omega^2z+(1-\tau z)(1-z)](R_1^2+R_2^2)+\nonumber \\
&&8\omega(1-\tau z^2)R_1R_2+
3[4\omega^2-(1-\tau z)^2]R_1^2R_2^2.
\label{eq_diffRzbar}
\end{eqnarray}
\end{widetext}
Now we may obtain an expression for the susceptibility as a function of the parameters $z$, $\tau$, and $\omega$, as well as of the ratios $R_1$ and $R_2$ and their derivatives with respect to ${\bar z}$. The expression is too long to be given here, but if we particularize it to the {\bf P} phase, where $R_1=R_2=R$, with $R$ given by Eq.~(\ref{rfp}) it simplifies to:
\begin{equation}
\chi_N=\frac{2(1+2\omega-\tau z)(1-z-2\omega z)}
{[1-2(1+2\omega)z+\tau z^2][1-(1+\tau-4\omega^2)z+\tau z^2]}.
\label{chi}
\end{equation}
At the {\bf P}-{\bf AN} transition $\tau_{critical}=1/z-4\omega^2/(z-1)$, as expected the denominator in Eq.~(\ref{chi}) vanishes, and thus we may write this equation as:
\begin{equation}
\chi_N=\frac{2(1+2\omega-\tau z)(1-z-2\omega z)}
{z(z-1)[1-2(1+2\omega)z+\tau z^2](\tau-\tau_{critical})}.
\label{chi1}
\end{equation}
We thus conclude that, in agreement with the findings of Fisher and Berker \cite{fb82} for discontinuous critical transitions, the transition from the regular polymerized to the nematic phase is characterized by the critical exponents $\beta=0$, since the nematic order parameter is discontinuous, and $\gamma=1$, as is clear in Eq.~(\ref{chi1}). Of course, the divergence may also be seen if we cross the critical line in another direction, such as parallel to the axis which represents the bond activity $z$, as long as we really cross it and do not touch the critical line tangentially. In Fig.~(\ref{chiws}) some of these curves are shown close to the transition activity.

\section{Final discussions and Conclusion}
\label{conc}

Semi-flexible trails on a Bethe lattice with coordination number equal to 4 show a very rich phase diagram in the parameter space defined by the activity of a bond ($z$), the statistical weights of a crossing and a collision ($\tau_c$ and $\tau_x$), and the statistical weight of an elementary bend in the trail ($\omega$). For sufficiently flexible chains ($\omega>1/2$) the phase diagrams are qualitatively similar to the one found in the flexible case ($\omega=1$), studied in \cite{tj16}, with non-polymerized ({\bf NP}), regular polymerized ({\bf P}) and dense polymerized ({\bf DP}) phases meeting at a bicritical point. When the Boltzmann factor of bends is smaller than 1/2, an additional polymerized phase appears inside the {\bf P} phase. In this phase all lattice sites are visited by the trails and all bonds are in one of the two possible directions, thus characterizing it as anisotropic and nematic ({\bf AN}). The nature of the {\bf P}-{\bf AN} transition is quite unusual: while the nematic order parameter is discontinuous, it also has a critical nature, characterized by the fact that the stability limits of both phases coincide with the transition line.  This type of criticality was studied in the framework of the renormalization group by Fisher and Berker \cite{fb82}, and for the present case we have verified their result that the susceptibility critical exponent $\gamma$ should be equal to 1. In the limit of rigid chains $\omega=0$ the {\bf P} phase is no longer stable, and three coexistence lines ({\bf AN}-{\bf NP}, {\bf AN}-{\bf DP}, and {\bf NP}-{\bf DP}) meet at a triple point which is the endpoint of the bicritical line.

Some features of the results presented here may be due to the particular lattice on which the model is solved. On the Bethe lattice, since there are no closed paths, any collision may be replaced by a crossing and vice-versa, so that the two statistical weights associated to these configurations only appear in the combination $\tau_x+2\tau_c$; this should no longer be true on a lattice with closed paths. Also, the phases {\bf DP} and {\bf AN} are totally frozen in the Bethe lattice solution: all lattice edges are occupied by bonds in the former, the same happening for all edges in one of the two directions in the latter. This also should change on lattices which are closer to regular ones. It is thus of interest to study this problem on the Husimi lattice, where small loops are present, and we are presently working on this.

\begin{figure}[t]
\centering
\includegraphics[width=7.0cm]{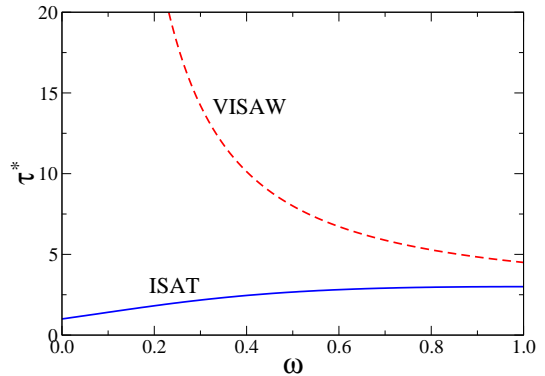}
\caption{Bicritical lines in terms of (on-site) monomer-monomer interaction $\tau^c*$ against $\omega$ for ISAT and VISAW models.}
\label{ISATxVISAW}
\end{figure}

It is interesting to compare the behavior of the ISAT model, where we choose $\tau_x=\tau^*$ and $\tau_c = \omega^2 \tau^*$, and the VISAW model, for which $\tau_x=0$ and $\tau_c = \omega^2 \tau^*$. In the former, the bicritical line is located at $\tau^*_{BC}=(1+2\omega)^2/(1+2\omega^2)$, while for the VISAW model it is at $\tau^*_{BC}=(1+2\omega)^2/2\omega^2$. These behaviors are compared in Fig.~\ref{ISATxVISAW}. In the ISAT model, as the stiffness of the chains is increased, the collapse transition becomes easier, since $\tau^*$ decreases as $\omega$ decreases. For the VISAW model, we notice an opposite behavior, with $\tau^* \rightarrow \infty$ when $\omega \rightarrow 0$, so that an infinite (on-site) monomer-monomer interaction is needed to collapse the chains. This is expected, since crossings are forbidden in VISAW and, thus, the stiffness will make the collapse more difficult. Moreover, this suggests that the globule phase for VISAW is similar to the one in the SASAW model. On the other hand, the results for ISAT's show that its collapsed phase is quite different from the one found in previous models. Again we could expect that such behavior should change on lattices with closed paths, as suggested by the results for the semi-flexible VISAW on the square lattice \cite{v15}.

\acknowledgments
We acknowledge support of the brazilian agencies CNPq and FAPEMIG, particularly from the former through project PVE 401228/2014-2, which made TP's stay in Brazil possible. We thank Profs.~M\'{a}rio J.~de Oliveira and Ron Dickman for helpful discussions regarding discontinuous critical transitions. TP acknowledges support by EPSRC grant EP/L026708/1. JFS thanks the Queen Mary University of London for hospitality.

\appendix*

\section{Rods ($\omega=0$) on the square lattice}

\begin{figure}[b]
\centering
\includegraphics[width=8.0cm]{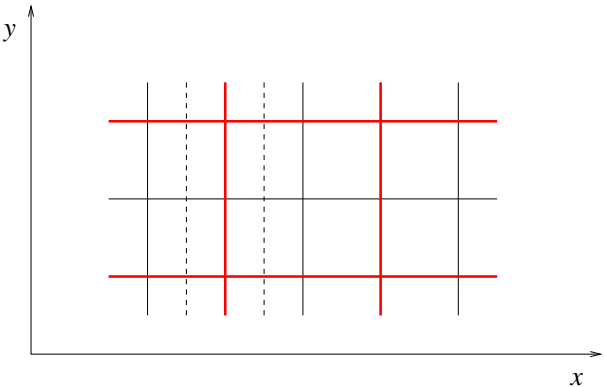}
\caption{A possible configuration of one vertical and two horizontal rods on a strip with $L_x=5$ and $L_y=3$. The element of the transfer matrix corresponding to the two successive $(1,0,1)$ configurations of horizontal edges indicated by the dashed lines is $z_1^2z_2^3\tau_x^2$.}
\label{strip}
\end{figure}

The particular case of the model where bends are forbidden ($\omega=0$) and therefore there are also no collisions (since $\tau_c=0$) as well, allows for a simple solution on the square lattice, which is a generalization of the case with no crossings discussed in \cite{sr15}. We may use a transfer matrix calculation to calculate the free energy, although this may also be done using combinatorial arguments. We start defining the model on a strip of the square lattice in the $(x,y)$ plane with finite width $L_y$ and length $L_x$, as shown in Fig.~\ref{strip}. Boundary conditions are periodic in both directions.

The states of the transfer matrix ${\mathbf T}$ will be defined by the set of $L_y$ horizontal edges between two successive vertical lines of the lattice. We therefore have $N_c=2^{L_y}$ possible configurations, which may be defined by a vector $|j\rangle=(\eta_{1,j},\eta_{2,j},\ldots,\eta_{L_y,j})$, where $\eta_{i,j}=0$ (1) if the edge $i$ is empty (occupied). The grand partition function will be
\begin{equation}
\Xi=Tr {\mathbf T}^{L_x}=\sum_{i=1}^{N_c} \lambda_i^{L_x},
\end{equation}
where $\lambda_i$ are the eigenvalues of the transfer matrix. In the thermodynamic limit $L_x \to \infty$, the partition function is dominated by the leading eigenvalue $\lambda_1$, so that the free energy per site will be:
\begin{equation}
\phi_b=-\lim_{L_x \to\infty}\frac{\ln \Xi}{L_x L_y}=
\frac{\ln \lambda_1}{L_y}.
\end{equation}
By construction, the transfer matrix is diagonal, and its eigenvalues are
\begin{equation}
\lambda_i={\mathbf T}_{i,i} = \langle i|{\mathbf T}|i\rangle = z_1^{s_i}+z_2^{L_y}(z_1 \tau_x)^{s_i},
\label{eqLambaTM}
\end{equation}
where $s_i=s_j=\sum_{k=1,L_y} \eta_k$ is the number of horizontal rods and $z_1$ and $z_2$ are the activities of horizontal and vertical rods, respectively. The two contributions in the right hand side of Eq.~(\ref{eqLambaTM}) correspond to having or not a vertical rod between the two horizontal edge sets, respectively. The number of states with $s_i$ rods is $L_y!/[s_i!(L_y-s_i)!]$. The functions $\lambda_i(s)$ are convex, so that the maxima should be located at $s=0$ or $s=L_y$. Now $\lambda(0)=1+z_2^{L_y}$, and $\lambda(L_y)=z_1^{L_y}+(z_1z_2\tau_x)^{L_y}$. The transition between these two phases occurs at $z_1^{L_y}-z_2^{L_y}+(z_1z_2\tau_x)^{L_y}=1$, which reduces to $z^2\tau_x=1$ in the symmetric case $z_1=z_2=z$. Thus, even in one dimension (finite values of $L_y$), a discontinuous phase transition happens between a phase {\bf A} with no horizontal rods ($\rho_1=0$) and a density
\begin{equation}
\rho_2=\frac{z_2}{L_y\lambda(0)}\frac{\partial \lambda(0)}{\partial z_2}=\frac{z_2^{L_y}}{1+z_2^{L_y}}
\end{equation}
for vertical ones, and a phase {\bf B} where all horizontal edges are occupied by bonds ($\rho_1=1$) and the density of vertical bonds is $\rho_2=(z_2\tau_x)^{L_y}/[1+(z_2\tau_x)^{L_y}]$.

In the two-dimensional limit $L_y \to \infty$, we find three phases, and the regions in the parameter space where each of them minimizes the free energy will be determined by $x=\max(1,z_1,z_2,z_1z_2\tau_x)$ 
\begin{itemize}
\item Non-polymerized {\bf NP}, where the densities of rods vanish ($\rho_1=\rho_2=0$), if $x=1$.
\item Anisotropic nematic ({\bf AN}), where all horizontal edges and no vertical edge is occupied ($\rho_1=1$ and $\rho_2=0$), if $x=z_1$; or all vertical and no horizontal edge is occupied ($\rho_1=0$ and $\rho_2=1$), if $x=z_2$.
\item A dense polymerized ({\bf DP}) phase, where all edges are occupied ($\rho_1=1$ and $\rho_2=1$), if $x=z_1z_2\tau_x$.
\end{itemize}
In the symmetric case, when $z_1=z_2=z$, the {\bf NP} phase is stable for $z \leqslant 1$, the {\bf AN} phase is stable in the region of the parameter space where $z\geqslant1$ and $\tau_x\leqslant 1/z$ and finally the stability region of the {\bf DP} phase is given by $\tau_x \geqslant 1/z$. Recalling that \textbf{NP} and \textbf{DP} phases came out from $\lambda_i(0)$ and $\lambda_i(L_y)$, respectively, they coexist at $\tau_x=1/z^2$. These results turn out to be identical to the ones for the Bethe lattice solution (in this particular case, with $\omega=0$), leading to the phase diagram obtained above and depicted in Fig.~\ref{pd}(d).

\end{document}